\documentclass[sigconf, nonacm]{acmart}

\usepackage{pvldb}

\usepackage{amsmath}

\usepackage{algorithm}
\usepackage{algorithmicx}
\usepackage[noend]{algpseudocode}

\usepackage{graphicx}
\usepackage[table]{xcolor}
\usepackage{enumitem}

\usepackage[hang,flushmargin]{footmisc}

\usepackage{caption}

\usepackage{booktabs}
\usepackage{tabularx}
\newcolumntype{Y}{>{\centering\arraybackslash}X}

\usepackage{url}
\newcommand{\sys}{\textsf{SieveIVF}}
\newcommand{\figcaption}[2]{%
  \caption{\textbf{#1} \textnormal{\textit{#2}}}%
}
\definecolor{SieveCodeBlue}{HTML}{4477AA}
\definecolor{SieveCodeGreen}{HTML}{3F7D5B}
\definecolor{SieveCodeGray}{HTML}{8B9298}
\definecolor{SieveCodeNote}{HTML}{696969}
\definecolor{SieveTableGray}{HTML}{F0F0F0}
\definecolor{SieveTableRule}{HTML}{555555}
\newcommand{\AlgKeyword}[1]{\textcolor{SieveCodeBlue}{\textbf{#1}}}
\newcommand{\AlgNote}[1]{\textcolor{SieveCodeNote}{\textit{// #1}}}
\algrenewcommand\algorithmicrequire{\AlgKeyword{Require:}}
\algrenewcommand\algorithmicwhile{\AlgKeyword{while}}
\algrenewcommand\algorithmicdo{\AlgKeyword{do}}
\algrenewcommand\algorithmicfor{\AlgKeyword{for}}
\algrenewcommand\algorithmicif{\AlgKeyword{if}}
\algrenewcommand\algorithmicthen{\AlgKeyword{then}}
\algrenewcommand\algorithmicelse{\AlgKeyword{else}}
\algrenewcommand\algorithmicprocedure{\AlgKeyword{procedure}}
\algrenewcommand\algorithmicreturn{\AlgKeyword{return}}
\algrenewcommand\textproc[1]{\textcolor{SieveCodeBlue}{\textsf{#1}}}
\algrenewcommand\algorithmiccomment[1]{%
  \hfill\textcolor{SieveCodeGreen}{\(\triangleright\)\ #1}}
\algrenewcommand\alglinenumber[1]{\textcolor{SieveCodeGray}{#1}}
\newenvironment{SievePseudocode}{\begingroup\small}{\endgroup}
\renewcommand\vldbdoi{XX.XX/XXX.XX}
\renewcommand\vldbpages{XXX-XXX}
\renewcommand\vldbavailabilityurl{}

\begin{document}

\title{SieveIVF: Threshold-Aware IVF Execution for Large-Scale Training Data Deduplication}

\author{%
  Zhisheng Hu$^{1,2}$,
  Zhifang Li$^{2}$,
  Junjie Chen$^{2}$,
  Ke Xu$^{2}$,
  Yuxuan Li$^{2}$, 
  Chufeng Chen$^{2}$, \\
  Rui Chen$^{2}$,
  Zhe Chen$^{2}$,
  and Ming-Chang Yang$^{1}$ \\
  $^{1}$\textit{The Chinese University of Hong Kong}
  \qquad
  $^{2}$\textit{Tencent}
}

\renewcommand{\vldbauthors}{
  Zhisheng Hu, Zhifang Li, Junjie Chen, Ke Xu, Yuxuan Li, Chufeng Chen, Rui Chen, Zhe Chen, and Ming-Chang Yang}

\renewcommand{\shortauthors}{Zhisheng Hu et al.}

\begin{abstract}
Embedding-based training data deduplication retrieves candidate duplicate edges above an application similarity threshold, but fixed-probe inverted-file (IVF) search ignores this predicate when giving every query the same partition budget. Across four Hunyuan workloads, qualifying neighbors appear early despite sharply varying search depths. We present \sys{}, a threshold-aware IVF executor that stops after $W$ consecutive searches find no qualifying candidate. The systems challenge is to preserve partition-major batching when each query's remaining work depends on prior results. Continuous batching groups ready queries by partition. A lookahead scheduler layers on top, exposing only work committed by the stopping rule to increase concurrency without changing stopping decisions or returned results. We implement \sys{} in Lance. At $W=8$, \sys{} is $4.1$--$7.6\times$ faster than fixed-probe IVF on four 10M Hunyuan workloads and $6.1$--$8.4\times$ faster on two public 100M workloads under the same index and search parameters, with pooled filtered top-10 recall losses of 0.03--1.13 percentage points on Hunyuan and 1.43--2.29 percentage points on the public workloads. These results show how an application predicate can guide IVF work allocation without changing the index or bounded top-$k$ interface.
\end{abstract}


\maketitle


\section{Introduction}\label{section_1_intro}

Large multimodal training corpora contain repeated screenshots, templated webpages, lightly edited images, and semantically redundant examples. Exact hashes identify byte-level duplicates but miss many of these relationships. Embedding-based training data deduplication instead retrieves similar records and uses qualifying neighbors as candidate edges for duplicate grouping. Prior work shows that removing such semantic duplicates can improve data and compute efficiency during model training~\cite{semdedup}.

Candidate retrieval for deduplication differs from generic top-$k$ search because it has an application similarity threshold $\tau$. A neighbor below $\tau$ cannot become a duplicate edge, regardless of its rank. Inverted-file (IVF) search, however, typically ranks partitions by centroid distance and probes a fixed number, $nprobe$, for every query before filtering the returned candidates. This gives easy and difficult queries the same search budget. Lowering $nprobe$ saves work for queries whose qualifying neighbors appear early, but it can truncate queries that require deeper probing. No single fixed probe budget accommodates both query types efficiently.

We study this mismatch using exact ground truth and stored IVF assignments from four 10M workloads in the Tencent Hunyuan training-data pipeline. Threshold-neighbor density varies sharply across queries and workloads, yet qualifying top-10 neighbors generally concentrate near the front of the partition order. The threshold can therefore guide search work instead of filtering output after every query exhausts the same fixed budget.

We present \sys{}, a threshold-aware IVF executor that follows the original partition order and stops each query after $W$ consecutive partition searches return no threshold-qualified candidate. The rule requires neither a learned predictor nor a new index. The consecutive-empty window $W$ is its only execution knob. Decreasing $W$ prunes more aggressively, whereas increasing it makes the rule more conservative and closer to fixed probing.

The central systems challenge is to retain efficient partition-major batching. Conventional batched IVF determines before execution which queries will search each partition. With data-dependent stopping, this grouping changes as individual queries finish. Executing each query independently preserves the rule but produces small partition searches, while dispatching every remaining partition restores concurrency but performs work beyond an uncommitted stopping point. \sys{} resolves this tension with \emph{continuous batching}, which maintains per-query state and dynamically groups ready queries by partition. Furthermore, a \emph{lookahead scheduler} layers on top of continuous batching to expose searches already committed by the stopping rule, increasing concurrency without changing the visited partitions, stopping decisions, or returned results.

We implement \sys{} in Lance~\cite{lance} without changing the IVF index format, within-partition search, or bounded top-$k$ interface. At $W=8$, \sys{} is $4.1$--$7.6\times$ faster than fixed-probe IVF on four 10M Hunyuan workloads and $6.1$--$8.4\times$ faster on two public 100M workloads under the same index and search parameters, with pooled filtered top-10 recall losses of 0.03--1.13 percentage points on Hunyuan and 1.43--2.29 percentage points on the public workloads.

\begin{samepage}
This paper makes the following contributions:
\begin{itemize}[leftmargin=*, itemsep=1pt, topsep=2pt]
    \item We characterize threshold selectivity and IVF locality across four Hunyuan workloads, then turn consecutive threshold-empty searches into a lightweight per-query stopping signal.
    \item We design continuous batching to preserve partition-major execution under data-dependent per-query stopping. A lookahead scheduler layers on top to expose additional committed work without changing the visited partitions or returned results.
    \item We implement \sys{} in Lance and evaluate it on four 10M Hunyuan workloads and two public 100M workloads. We compare a shared fixed-probe budget with workload-tuned baselines and isolate the effects of pruning, batching, and lookahead.
\end{itemize}
\end{samepage}

\section{Background}\label{section_background}

\subsection{Training Data Deduplication}

Embedding-based training data deduplication pipelines first convert raw training records into vectors, then query a vector index to retrieve near-duplicate candidates above a similarity threshold. Candidate edges may then be filtered using application-specific content signals, such as OCR and associated QA text. The surviving candidate edges are then grouped into duplicate components, which are used to assign labels for downstream training. Our production context is a Tencent Hunyuan data pipeline that prepares deduplicated corpora for large-scale multimodal model pretraining. Duplicate semantics remain application-specific. For example, SemDeDup defines semantic redundancy in embedding space and studies how removing redundant records affects training~\cite{semdedup}. Once an application has selected such a policy, including its embedding model and similarity threshold, \sys{} accelerates only candidate retrieval and leaves the policy itself unchanged.

\subsection{Threshold Candidate Retrieval}

Let $X=\{x_1,\ldots,x_N\}$ be the indexed corpus and $Q$ be a batch of query embeddings. Given a similarity threshold $\tau$, the natural deduplication candidate set for query $q$ is
\[
    R(q,\tau)=\{x \in X \mid \mathrm{sim}(q,x)\ge \tau\}.
\]
Vector search APIs often return at most $k$ records per query, so the system may compute a top-$k$ subset of $R(q,\tau)$. The threshold $\tau$ determines which candidates qualify as potential duplicates. \sys{} uses the presence or absence of such candidates in each partition to decide how much IVF work to perform, regardless of how the underlying search engine applies the threshold internally.

\subsection{IVF Batch Search}

Inverted file (IVF) indexes are widely used to scale vector search~\cite{jegou2011product,spann,milvus}. An IVF index clusters the corpus into partitions represented by centroids. For a query $q$, the search engine ranks centroids by distance to $q$, probes the nearest $nprobe$ partitions, performs within-partition search~\cite{quickadc}, and merges candidate results. Training data deduplication processes large query batches offline, so our executor uses partition-major batching to improve throughput. It groups queries that probe the same partition so that they can share partition loading, cache locality, and CPU execution.

This batch structure amortizes each partition's fixed search cost across many queries, so throughput depends on keeping queries grouped by partition. If an optimization turns one large batch into many small per-query calls, it may reduce the number of searched partitions but lose the locality that makes IVF efficient. \sys{} is designed around this constraint: it introduces per-query early stopping while preserving partition-major batching.

\section{Workload Characterization}\label{section_motivation}

\subsection{Hunyuan Workloads and Methodology}

We study four proprietary training-data deduplication workloads from the Tencent Hunyuan data pipeline: Webpage (WEB), Table-QA (TABLE), STEM-Webpage (STEM), and Scene-OCR-QA (SCENE). Each starts with 10M 768-dimensional image embeddings. We uniformly sample 100K queries and remove them from the base, leaving an exclusive 9.9M-vector base. The workloads share the embedding representation but apply different downstream policies and cosine-similarity thresholds: 0.85 for WEB, 0.93 for TABLE, 0.90 for STEM, and 0.93 for SCENE. Each is calibrated per workload via offline sampling analysis before deduplication.

Each 9.9M-vector base uses an IVF index with 2,400 partitions and RaBitQ~\cite{rabitq} configured with 7 bits, which we denote as IVF\_RQ7. The partition count gives approximately 4,125 vectors per partition, close to Lance's recommended target of 4,096~\cite{lance}. We derive two workload measurements from exact ground truth. To measure threshold-neighborhood density, we exhaustively compute each query's exact top-50 under cosine similarity, matching the largest $k$ evaluated later, and count the neighbors that meet its threshold. To measure IVF locality, we take the threshold-filtered exact top-10 neighbors, locate their stored partitions, and record where those partitions occur in the probe order used by Lance's cosine search. All indexes use exact centroid assignment. Section~\ref{section_eval_assignment} evaluates approximate assignment separately.

\subsection{Threshold Selectivity Varies Sharply}

\begin{figure}[t]
    \centering
    \includegraphics[width=3.34in]{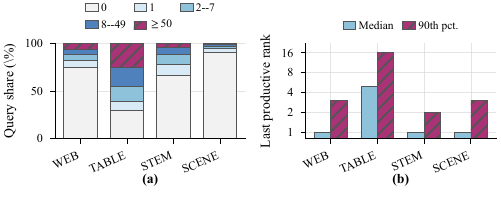}
    \figcaption{Exact-GT characterization of Hunyuan workloads.}{}
    \label{fig:workload-characterization}
    \Description{A stacked bar chart shows threshold-neighbor density across four Hunyuan workloads. A grouped bar chart shows the median and 90th-percentile rank of the last IVF partition containing a filtered exact top-10 neighbor.}
\end{figure}

Figure~\ref{fig:workload-characterization}(a) shows that these workloads occupy substantially different selectivity regimes. SCENE is sparse: 90.77\% of queries have no threshold neighbor. WEB and STEM have zero-neighbor shares of 75.09\% and 66.94\%. TABLE is much denser: only 29.51\% have no threshold neighbor, while 25.36\% reach the right-censored $\geq 50$ bin. Consequently, a production pipeline cannot treat one modality's duplicate density as representative of the others.

This heterogeneity exposes the limitation of a query-agnostic probe budget. Fixed-probe IVF assigns the same $nprobe$ to every query. A value large enough for dense, long-tailed queries wastes work on sparse queries, while a smaller value may miss useful neighbors that appear later. Density variation therefore motivates a per-query probe budget, but does not by itself justify early stopping. Empty partitions are informative only when threshold-qualified neighbors typically occur near the front of the probe order.

\subsection{Exact Neighbors Are Front-Loaded}

Figure~\ref{fig:workload-characterization}(b) reports the rank of the last IVF partition containing a threshold-filtered exact top-10 neighbor. For WEB, STEM, and SCENE, the median last rank is 1 and the 90th percentile is only 2--3, whereas TABLE reaches 5 and 16, respectively.

Pooling threshold-qualified exact top-10 neighbors across queries, the first eight centroid-ranked partitions store 99.77\%, 99.95\%, and 99.87\% of these neighbors for WEB, STEM, and SCENE, respectively. TABLE reaches 93.76\% at rank 8 and 98.31\% at rank 16. These values measure partition placement rather than search recall because compressed-vector search introduces additional approximation. They show that useful neighbors are strongly front-loaded, but the tail length differs across workloads. TABLE therefore provides a difficult case for a common stopping policy.

\subsection{The Opportunity Is Semantic, Not Physical}

The preceding sparsity is query dependent; it is not caused by an accidentally empty index. Table~\ref{tab:partition-occupancy} shows that every one of the 9,600 physical partitions across the four indexes contains vectors. Partition sizes have coefficients of variation from 0.590 to 0.755, and the largest partition is 5.61--7.14$\times$ the mean. Thus the indexes exhibit real, nonuniform occupancy but no empty-partition shortcut. TABLE is the most balanced index by both coefficient of variation and Gini coefficient, yet has the most diffuse qualifying neighbors. The opportunity therefore comes from \emph{threshold-empty results for a particular query}, not from skipping physically empty partitions.

\begin{table}[t]
    \centering
    \figcaption{Physical IVF partition-size statistics.}{}
    \label{tab:partition-occupancy}
    \small
    \renewcommand{\arraystretch}{0.92}
    \setlength{\tabcolsep}{2pt}
    \setlength{\arrayrulewidth}{0.35pt}
    \setlength{\aboverulesep}{0.6pt}
    \setlength{\belowrulesep}{0.6pt}
    \arrayrulecolor{SieveTableRule}
    \rowcolors{2}{SieveTableGray}{white}
    \begin{tabularx}{\columnwidth}{Y|Y|Y|Y}
        \toprule
        Workload & CV & Gini & Max/mean \\
        \midrule
        WEB & 0.755 & 0.329 & 6.44$\times$ \\
        TABLE & 0.590 & 0.240 & 5.61$\times$ \\
        STEM & 0.721 & 0.316 & 7.14$\times$ \\
        SCENE & 0.664 & 0.305 & 6.35$\times$ \\
        \bottomrule
    \end{tabularx}
\end{table}

\subsection{From Workload Evidence to a Search Policy}

The characterization exposes a limitation of fixed-probe IVF. It visits the same number of partitions for every query even though useful-neighbor density and span vary sharply across queries and workloads. Simply lowering $nprobe$ removes the same suffix for all queries and risks truncating TABLE-like queries whose useful neighbors extend farther into the probe order. Changing the similarity threshold instead changes the deduplication task. What is missing is a runtime signal that adapts each query's executed probe count without changing $\tau$ or its maximum $nprobe$ budget.

\begin{figure}[t]
    \centering
    \includegraphics[width=\linewidth]{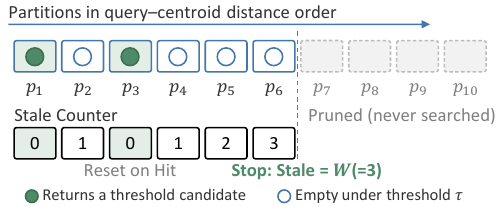}
    \figcaption{Threshold-aware stopping over IVF partitions.}{}
    \label{fig:stopping}
    \Description{A query visits IVF partitions in centroid-distance order. Hits reset a stale counter, empty results increment it, and three consecutive empty results prune the remaining partitions.}
\end{figure}

The local search already provides such an observation. A partition is \emph{threshold-empty} for a query when its local search returns no candidate satisfying $\tau$. A single empty result is inconclusive: IVF search is approximate and a later partition may still qualify. A run of empty results, however, is runtime evidence that the query has left the front-loaded productive region observed above. \sys{} therefore uses a consecutive-empty window $W$. It stops a query when the current run reaches $W$ threshold-empty partitions (Figure~\ref{fig:stopping}). A hit resets the counter, allowing a dense or long-tailed query to continue. A small $W$ reacts quickly and saves more work; a large $W$ is conservative and approaches fixed probing.

The consecutive-run form matters. Unlike a cumulative empty count or average empty rate, it treats any later hit as evidence that the query remains in a productive region. Between hits the counter is monotone, so the partitions that must be searched next can be committed solely from completed observations. Section~\ref{section_design} uses this property to make batched execution reproduce the same sequential stopping rule without predicting future hits.

\subsection{The Batch Execution Challenge}

A useful stopping rule is not yet an efficient operator. Executing it independently for each query fragments IVF into tiny partition searches and loses the shared loading, cache locality, and CPU efficiency of partition-major batching. At the other extreme, dispatching a query's entire remaining probe list restores concurrency but schedules ranks before completed results have committed them. Such speculative work weakens pruning and can change the rule's output if results beyond the eventual stopping point are retained.

The execution design must therefore reduce per-query work, preserve partition-local batching as queries reach different stopping points, and expose enough committed work when few queries are active. Section~\ref{section_design} addresses these requirements with continuous batching as the execution substrate and a lookahead scheduler layered on top. Continuous batching dynamically regroups active queries by partition. Lookahead adds concurrency by exposing only ranks already committed by the same stopping rule.

\section{\sys{} Design}\label{section_design}

The dilemma in Section~\ref{section_motivation} yields four requirements. \textbf{R1 (compatibility):} run on an unmodified IVF index with the same centroids, partitions, and within-partition search logic, so the technique is a drop-in execution policy rather than a new index. \textbf{R2 (batch-preserving):} reduce per-query work without destroying the partition locality that batched IVF depends on. \textbf{R3 (concurrency-robust):} serve both large offline batches and low-concurrency calls. \textbf{R4 (minimal interface):} expose one execution knob rather than another application policy, because deduplication pipelines already tune embedding, OCR, and text thresholds. The stopping rule below satisfies R1 and R4 directly. Continuous batching (Section~\ref{section_design_continuous}) addresses R2 and the many-query case of R3, and lookahead scheduling (Section~\ref{section_design_lookahead}) extends R3 to low-concurrency calls.

\subsection{Threshold-Aware Ordered Traversal}\label{section_design_traversal}

For each query $q_i$ in a batch, IVF ranks candidate partitions by query-centroid distance and produces the ordered list:
\[
P_i = (p_{i,1}, p_{i,2}, \ldots, p_{i,nprobe}).
\]
\sys{} preserves this order and changes only how far each query proceeds through the list before stopping.

Each query maintains the state in Table~\ref{tab:query-state}.
Here, a partition is \emph{empty} for a query if the local search returns no candidate satisfying $\tau$, not if the partition contains no indexed vectors. After a partition search, if the result is non-empty, \sys{} appends it to \texttt{results} and resets \texttt{stale} to zero. If the result is empty, \texttt{stale} is incremented. Once \texttt{stale} reaches $W$, the query is stopped.

\begin{table}[t]
    \centering
    \figcaption{Per-query state under continuous batching.}{Lookahead instead uses stop line $s_i$ and enqueue cursor $e_i$ (Section~\ref{section_design_lookahead}).}
    \label{tab:query-state}
    \small
    \renewcommand{\arraystretch}{0.92}
    \setlength{\tabcolsep}{2pt}
    \setlength{\arrayrulewidth}{0.35pt}
    \setlength{\aboverulesep}{0.6pt}
    \setlength{\belowrulesep}{0.6pt}
    \arrayrulecolor{SieveTableRule}
    \rowcolors{2}{SieveTableGray}{white}
    \begin{tabularx}{\columnwidth}{>{\centering\arraybackslash}m{0.22\columnwidth}|Y}
        \toprule
        State & Meaning \\
        \midrule
        \texttt{next} & Next rank in $P_i$ to process \\
        \texttt{stale} & Consecutive threshold-empty result count \\
        \texttt{results} & Accumulated non-empty local results \\
        \bottomrule
    \end{tabularx}
\end{table}

\subsection{Continuous Batching}\label{section_design_continuous}

\begin{figure}[t]
    \centering
    \includegraphics[width=\linewidth]{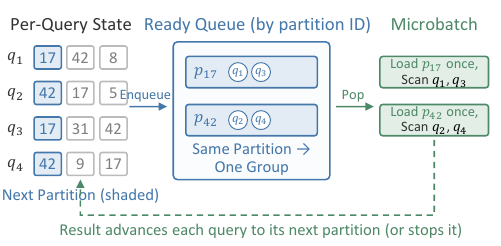}
    \figcaption{Continuous batching groups queries by partition.}{}
    \label{fig:continuous}
    \Description{Four queries expose their next partition. Queries sharing partition 17 or 42 are grouped in a ready queue and searched as partition-local microbatches before their state advances.}
\end{figure}

Fixed-probe batching can form all partition batches in advance. With per-query stopping, however, whether a query should search its next partition is known only after earlier searches complete. \sys{} therefore maintains per-query execution state and continuously groups queries that are ready to search the same IVF partition (Figure~\ref{fig:continuous}, Algorithm~\ref{alg:continuous-pruning}).

\noindent\textbf{Dynamic Regrouping.}
Each \emph{ready item} $(q_i,j)$ indicates that query $q_i$ is ready to search partition $p_{i,j}$. Items targeting the same IVF partition form a \emph{partition group}. A \emph{microbatch} is a bounded subset of one group that workers search together. Per-query stopping makes group membership data-dependent because each completed search determines whether its query contributes another item. Continuous batching therefore updates the groups online from currently ready items, preserving partition locality as queries stop at different ranks.

\noindent\textbf{Scheduling Partition Groups.}
The ready queue processes partition groups in first-in, first-out (FIFO) order. A group enters the queue when it receives its first item, and later items for the same partition join it until dispatch. Each queue pop selects a target partition $p$ and up to 16K ready items from its group, forming a microbatch $B$. Larger groups are therefore served across multiple dispatches. FIFO determines only which ready group is served next; the key distinction from fixed-probe batching is online regrouping under data-dependent membership. During development, we also experiment with random and largest-group-first ordering, but neither consistently outperforms FIFO. We therefore retain FIFO for its simplicity and competitive performance.

Algorithm~\ref{alg:continuous-pruning} initializes one ready item per query at rank 1. Each completion updates the query's local results and \texttt{stale} counter, then enqueues the next rank if the query remains active (lines~\ref{line:cb-completion-begin}--\ref{line:cb-enqueue-next}). After all searches complete, lines~\ref{line:cb-final-merge-begin}--\ref{line:cb-return} merge each query's local results into its final top-$k$ output. A query thus contributes at most one ready item at a time and preserves the order in $P_i$. With few active queries, however, this progression can limit parallelism. Section~\ref{section_design_lookahead} addresses that case with lookahead scheduling.

\begin{algorithm}[t]
\caption{Continuous batching with per-query stopping}
\label{alg:continuous-pruning}
\begin{SievePseudocode}
\begin{algorithmic}[1]
\State \AlgNote{$Q$: query batch}
\State \AlgNote{$P=(P_i)_{q_i\in Q}$: per-query centroid-ranked partition orders}
\State \AlgNote{$\tau$: application similarity threshold; $W$: consecutive-empty window; $k$: neighbors per query}
\Procedure{ContinuousBatching}{$Q,P,\tau,W,k$}
    \State $(next_i,stale_i,results_i) \gets (1,0,\emptyset)$ for each $q_i \in Q$
    \State initialize ready queue with $(q_i,1)$ for each $q_i \in Q$ \label{line:cb-initial-admission}
    \While{ready or in-flight work remains}
        \While{worker available and ready queue non-empty}
            \State $(p,B) \gets$ \Call{PopMicrobatch}{ready queue}
            \State \Call{LaunchSearch}{$p,B,\tau,k$}
        \EndWhile
        \For{each completion $(q_i,j,r)$} \label{line:cb-completion-begin} \Comment{$j$: rank; $r$: search result}
            \State $next_i \gets j + 1$
            \If{$r \ne \emptyset$}
                \State append $r$ to $results_i$; $stale_i \gets 0$ \Comment{hit}
            \Else
                \State $stale_i \gets stale_i + 1$ \Comment{empty}
            \EndIf
            \If{$stale_i < W$ and $next_i \le |P_i|$} \label{line:cb-continuation-begin}
                \State enqueue $(q_i,next_i)$ \label{line:cb-enqueue-next}
            \EndIf
        \EndFor
    \EndWhile
    \For{each $q_i \in Q$} \label{line:cb-final-merge-begin}
        \State $R_i \gets$ \Call{MergeTopK}{$results_i,k$}
    \EndFor
    \State \Return $R=(R_i)_{q_i\in Q}$ \label{line:cb-return}
\EndProcedure
\end{algorithmic}
\end{SievePseudocode}
\end{algorithm}

\subsection{Lookahead Scheduling}\label{section_design_lookahead}

\noindent\textbf{Low Concurrency.}
Continuous batching gains throughput by merging queries that are simultaneously ready for the same partition. This works well when many queries are active, because partition groups can accumulate several ready items before dispatch. With only a handful of queries, however, a group often contains one item. Each query also exposes at most one ready item and waits for its result before advancing. Fixed-probe search commits to all $nprobe$ partitions up front and can overlap them without this dependency, so serialization and queue bookkeeping can outweigh the work saved by pruning under low concurrency.

\begin{figure}[t]
    \centering
    \includegraphics[width=\linewidth]{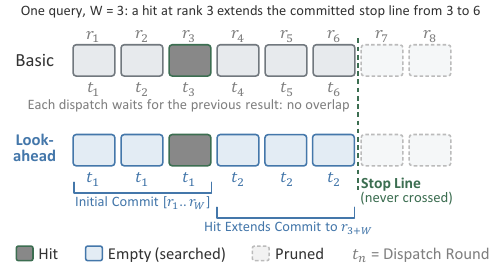}
    \figcaption{Lookahead advances dispatch within the committed prefix.}{Basic denotes continuous batching without lookahead.}
    \label{fig:lookahead}
    \Description{For one query with W set to three, continuous batching without lookahead searches committed ranks sequentially, while lookahead dispatches the same ranks in parallel rounds. A hit at rank three extends the stop line to rank six in both configurations.}
    \vspace{4pt}
\end{figure}

Simply dispatching a query's remaining tail would restore concurrency but would also schedule ranks whose necessity still depends on unseen hits. Lookahead instead uses work that the stopping rule has already committed. Every query must search ranks $1$ through $\min(W,nprobe)$, and every hit commits the next $W$ ranks. The committed ranks therefore form a prefix whose last rank is the query's \emph{stop line}. Lookahead dispatches this prefix without changing how far the query is allowed to travel (Figure~\ref{fig:lookahead}).

\noindent\textbf{Stop Line.}
Formally, a query $q_i$'s stop line $s_i$ is the largest partition rank such that $q_i$ is committed to search every partition up to $s_i$ before it may stop. Initially $s_i = \min(W,nprobe)$. The query must complete every rank through $s_i$ even if all of their results are empty. Whenever partition $p_{i,j}$ returns a hit, the consecutive-empty counter resets and the query commits the next $W$ ranks after $j$: $s_i \gets \max(s_i, \min(j+W,nprobe))$. Ranks at most $s_i$ are committed. Only later ranks depend on hits that have not yet happened.

Lookahead scheduling enqueues each committed rank as soon as it becomes committed, instead of waiting for the preceding partition's result. A single query can therefore contribute several ready items at once. Multiple low-concurrency queries are then more likely to contribute items to the same partition group and form a larger microbatch. Crucially, lookahead never expands the search beyond the stop line implied by the stopping rule. It changes \emph{when} committed partition searches are dispatched, not \emph{which} ranks the pruning policy is allowed to reach.

Algorithm~\ref{alg:lookahead} shows the resulting scheduler. It reuses Algorithm~\ref{alg:continuous-pruning}'s partition queue, worker dispatch, local search, per-query result buffers, and final merge, but replaces its initial admission (line~\ref{line:cb-initial-admission}) and per-query continuation logic (lines~\ref{line:cb-continuation-begin}--\ref{line:cb-enqueue-next}) with committed-prefix admission. Each query keeps $e_i$, the first rank not yet enqueued, in addition to the stop line $s_i$. Initialization enqueues ranks $1$ through $\min(W,nprobe)$ for each query. An empty result does not move $s_i$. A hit at rank $j$ commits through $\min(j+W,nprobe)$, after which the scheduler enqueues any newly committed suffix. Each rank is enqueued at most once, so lookahead adds parallelism without duplicating partition searches.

\begin{algorithm}[t]
\caption{Lookahead scheduling for one query}
\label{alg:lookahead}
\begin{SievePseudocode}
\begin{algorithmic}[1]
\State \AlgNote{$q_i$: query with ordered partitions $P_i$}
\State \AlgNote{$s_i$: stop line; $e_i$: first rank not yet enqueued}
\State \AlgNote{$W$: consecutive-empty window; $nprobe$: maximum partition rank}
\Procedure{EnqueueCommitted}{$q_i,s_i,e_i$}
    \While{$e_i \le s_i$}
        \State enqueue $(q_i,e_i)$
        \State $e_i \gets e_i + 1$
    \EndWhile
    \State \Return $e_i$
\EndProcedure
\Procedure{LookaheadSchedule}{$q_i,W,nprobe$}
	    \State $s_i \gets \min(W,nprobe)$; $e_i \gets 1$
	    \State $e_i \gets \Call{EnqueueCommitted}{q_i,s_i,e_i}$ \Comment{initial prefix}
	    \For{each completion $(q_i,j,r)$} \Comment{$j$: rank; $r$: search result}
	        \If{$r \ne \emptyset$}
	            \State $s_i \gets \max(s_i, \min(j+W,nprobe))$ \Comment{extend stop line}
	            \State $e_i \gets \Call{EnqueueCommitted}{q_i,s_i,e_i}$ \Comment{new suffix}
	        \EndIf
    \EndFor
\EndProcedure
\end{algorithmic}
\end{SievePseudocode}
\end{algorithm}

\noindent\textbf{Equivalence.}
Without lookahead, Algorithm~\ref{alg:continuous-pruning} advances a query only after its preceding partition search completes, so it realizes the sequential stopping rule. Lookahead initially admits ranks $1$ through $\min(W,nprobe)$ and, when a completed search at rank $j$ returns a hit, updates the stop line to $\max(s_i,\min(j+W,nprobe))$. A query--partition search's hit-or-empty outcome does not depend on when it is dispatched. Because stop-line extensions arise only from already committed ranks and combine through monotone maximum updates, concurrent completion order cannot change the final stop line. Every admitted rank is therefore one that the sequential rule is also committed to visit, while the monotone enqueue cursor admits every rank through that stop line exactly once. The two configurations consequently execute the same query--partition searches and merge the same local results. Lookahead changes dispatch timing, not stopping decisions or returned results.

\subsection{Discussion}

\noindent\textbf{Tradeoff and Precondition.}
\sys{} is a speed/recall optimization, not a lossless rewrite. A threshold-qualified candidate can appear after a run of empty partitions, and a small $W$ may stop before reaching it. The window therefore controls a real speed/recall tradeoff. \sys{} also supports online calibration that selects $W$ from a small sample of the same query batch under a baseline-relative loss target (\textbf{Appendix~\ref{section_appendix_calibration}}). For training-data deduplication, a modest loss in candidate recall can be acceptable because a missed edge leaves some redundancy, whereas an incorrect edge may remove distinct training data. Lookahead changes only dispatch timing and leaves this tradeoff unchanged. The stopping signal also assumes that search-time centroid order is consistent with how vectors were assigned to partitions at construction. \sys{} therefore requires exact centroid assignment. Section~\ref{section_eval_assignment} deliberately violates this build-time precondition to measure its effect.

\noindent\textbf{Memory Efficiency.}
\sys{} adds constant-sized state per query, and each ready item carries only query and partition metadata. Fixed-probe batching enumerates all $|Q|\,nprobe$ assignments before search, whereas continuous batching materializes only eligible work and lookahead only committed work. Early stopping therefore avoids scheduling metadata and local result batches for skipped searches, reducing memory relative to fixed probing.

\section{Implementation}\label{section_implementation}

We implement \sys{} in Lance~\cite{lance}, an open lakehouse format for multimodal AI with native vector search. The implementation reuses Lance's on-disk IVF layout, within-partition search logic, and final top-$k$ merge, and changes only how query--partition searches are admitted and scheduled at runtime.

Lance already accepts $\tau$ as a distance upper bound and applies it to filter each partition's local output. \sys{} consumes the resulting hit-or-empty outcome to update the per-query state that drives the stopping rule (Section~\ref{section_design_traversal}). The executor maintains these states together with the ready queue and partition groups. It forms a microbatch from up to 16K ready items for one partition and dispatches it to Lance's search worker pool. As each microbatch completes, the executor updates the affected queries and admits any newly committed ranks. This integration requires neither a new index nor a learned stopping model.

\section{Evaluation}\label{section_evaluation}

\subsection{Setup and Metrics}\label{section_eval_setup}

\noindent\textbf{Workloads.}
We evaluate the four proprietary 10M Hunyuan workloads characterized in Section~\ref{section_motivation} and two public 100M workloads. For each workload, we uniformly sample 100K vectors without replacement as held-out queries and index the remainder. Table~\ref{tab:evaluation-workloads} summarizes the resulting base sizes, dimensions, thresholds, and IVF layouts. Each Hunyuan workload uses the workload-specific threshold reported in Section~\ref{section_motivation}. LAION-100M uses embeddings from LAION-5B~\cite{laion5b}, while DEEP-100M uses descriptors from the first 100M vectors of DEEP1B~\cite{deep1b}. Both public workloads use a cosine-similarity threshold of $\tau=0.85$.

\begin{table}[t]
    \centering
    \figcaption{Evaluation workloads.}{IVF layout gives indexes $\times$ partitions per index. Data is raw FP32 base-vector size in decimal GB.}
    \label{tab:evaluation-workloads}
    \footnotesize
    \renewcommand{\arraystretch}{0.92}
    \setlength{\tabcolsep}{2pt}
    \setlength{\arrayrulewidth}{0.35pt}
    \setlength{\aboverulesep}{0.6pt}
    \setlength{\belowrulesep}{0.6pt}
    \arrayrulecolor{SieveTableRule}
    \rowcolors{2}{SieveTableGray}{white}
    \begin{tabularx}{\columnwidth}{
        >{\centering\arraybackslash}m{0.20\columnwidth}|
        >{\centering\arraybackslash}m{0.09\columnwidth}|
        >{\centering\arraybackslash}m{0.07\columnwidth}|
        >{\centering\arraybackslash}m{0.11\columnwidth}|
        >{\centering\arraybackslash}m{0.06\columnwidth}|
        >{\centering\arraybackslash}m{0.18\columnwidth}|
        Y}
        \toprule
        Workload & Base & Dim. & Queries & $\tau$ & IVF layout & Data (GB) \\
        \midrule
        WEB        & 9.9M  & 768 & 100K & 0.85 & $1\times2{,}400$  & 30.4 \\
        TABLE      & 9.9M  & 768 & 100K & 0.93 & $1\times2{,}400$  & 30.4 \\
        STEM       & 9.9M  & 768 & 100K & 0.90 & $1\times2{,}400$  & 30.4 \\
        SCENE      & 9.9M  & 768 & 100K & 0.93 & $1\times2{,}400$  & 30.4 \\
        LAION-100M & 99.9M & 512 & 100K & 0.85 & $16\times1{,}500$ & 204.6 \\
        DEEP-100M  & 99.9M & 96  & 100K & 0.85 & $16\times1{,}500$ & 38.4 \\
        \bottomrule
    \end{tabularx}
\end{table}

\noindent\textbf{Index and Measurement.}
All workloads use cosine IVF\_RQ7 indexes with exact centroid assignment. Lance normalizes base vectors at index time and queries at search time, and exact ground truth uses the same normalization. We choose IVF partition counts around Lance's recommended 4,096 vectors per partition~\cite{lance}. Each Hunyuan workload uses one index with 2,400 IVF partitions. Each public 100M workload has 16 coarse KMeans shards, with a separately built 1,500-partition Lance IVF\_RQ7 index for each shard.

Unless varied, fixed probing and \sys{} use the same index, $nprobe=90$, $k=10$, and threshold, while \sys{} uses $W=8$ with lookahead. Public queries search all 16 shards at $nprobe=90$ with identical four-way concurrency across executors, and their timing includes the global top-10 merge. Experiments use the Lance implementation described in Section~\ref{section_implementation} on a Tencent Cloud server with 32 logical cores on an AMD EPYC 7K62 CPU, 64\,GiB of memory, and two 1\,TB Tencent Cloud Block Storage (CBS) volumes of the \texttt{Cloud\,SSD} type. We report warmed-index wall-clock time under identical cache, warm-up, and concurrency conditions. Timing covers search and result collection, while setup and offline recall scoring remain outside every reported search time.

\noindent\textbf{Quality Metric.}
We compute ground truth by exhaustive cosine search over the full base, independently of IVF and pruning. Public experiments fix $k=10$ and use exact global top-10 lists. Let $G_{\tau,k}(q)$ be the threshold-filtered exact top-$k$ ground-truth set for $q$, and let $A(q)$ be the returned neighbor set. Aggregating over all queries gives our primary metric, \emph{pooled filtered top-$k$ recall}:
\[
R_{\mathrm{pool}}=
\frac{\sum_{q\in Q}|A(q)\cap G_{\tau,k}(q)|}
     {\sum_{q\in Q}|G_{\tau,k}(q)|}.
\]
Queries with empty $G_{\tau,k}(q)$ remain in the timed workload but do not contribute to the denominator. The DiskJoin comparison in Section~\ref{section_eval_alternative} reports direct-edge and connected-component coverage.

\subsection{Effectiveness and Scale}\label{section_eval_main}

\begin{figure}[t]
    \centering
    \includegraphics[width=3.34in]{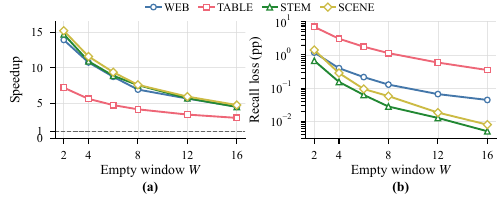}
    \figcaption{Speed/recall tradeoff across empty windows.}{}
    \label{fig:main-tradeoff}
    \Description{Two line charts show speedup and pooled filtered top-10 recall loss as the empty window W increases for WEB, TABLE, STEM, and SCENE.}
\end{figure}

Figure~\ref{fig:main-tradeoff} sweeps the consecutive-empty window while holding the index, $nprobe$, $k$, and threshold fixed. At the main setting $W=8$, \sys{} is $4.1$--$7.6\times$ faster than fixed probing with 0.03--1.13 percentage points of pooled filtered top-10 recall loss. The workload characterization in Section~\ref{section_motivation} explains this asymmetry. TABLE has the most qualifying neighbors, which extend furthest into the probe order, so it shows the smallest speedup and the largest recall change, while each of the other three workloads loses at most 0.13 percentage points of pooled filtered top-10 recall at $W=8$.

The full sweep shows why $W$ remains an explicit knob. At $W=2$, speedup rises to $7.2$--$15.2\times$, but TABLE loses 7.03 percentage points of pooled recall. At $W=16$, the loss falls to 0.005--0.351 percentage points while speedup remains $2.9$--$4.7\times$. No single window is optimal for every workload. We use $W=8$ in later experiments as a common operating point between these extremes.

\begin{figure}[t]
    \centering
    \includegraphics[width=3.34in]{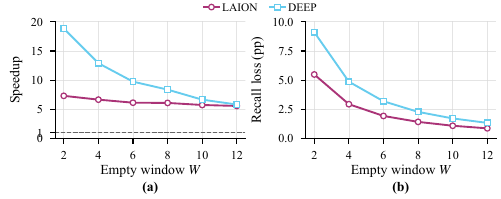}
    \figcaption{Scale-out results on two public 100M workloads.}{}
    \label{fig:public-100m-tradeoff}
    \Description{Two line charts show speedup and pooled filtered top-10 recall loss as the empty window W increases on LAION-100M and DEEP-100M.}
\end{figure}

Figure~\ref{fig:public-100m-tradeoff} repeats the sweep on the public workloads. At $W=8$, \sys{} is $6.11\times$ faster on LAION-100M and $8.38\times$ faster on DEEP-100M, with 1.43 and 2.29 percentage points of pooled recall loss. Increasing $W$ improves recall monotonically. At $W=12$, the two workloads retain $5.59\times$ and $5.83\times$ speedups while reducing the losses to 0.88 and 1.34 percentage points. These results extend the same execution mechanism to 100M-vector scale-out workloads with different dimensions and embedding geometries.

\begin{figure}[t]
    \centering
    \includegraphics[width=3.34in]{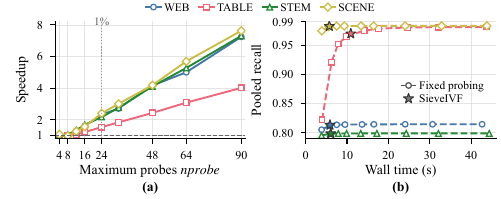}
    \figcaption{Probe-budget sensitivity and tuned fixed probing.}{}
    \label{fig:nprobe-fairness}
    \Description{Panel (a) plots same-nprobe speedup from nprobe 4 through 90 and marks nprobe 24 as 1 percent of the 2,400 IVF partitions. Panel (b) plots the fixed-probing recall-latency frontier for each Hunyuan workload and marks the SieveIVF result at W equals 8 and nprobe equals 90.}
    \vspace{-3pt}
\end{figure}

\noindent\textbf{Tuned Fixed-Probe Baseline.}
In deployment, $nprobe$ is set by index scale rather than by a per-workload ground-truth study, so all four workloads use 90 of 2,400 partitions. No shared fixed budget is efficient across all four. WEB, STEM, and SCENE saturate early, while TABLE benefits from deeper probing (Section~\ref{section_motivation}). \sys{} instead adapts per-query work with one shared $W=8$. With both executors capped at $nprobe=24$ (1\% of the 2,400 partitions), Figure~\ref{fig:nprobe-fairness}(a) shows that \sys{} remains $1.53$--$2.39\times$ faster with only 0.02--0.72 percentage points of pooled recall loss. Panel (b) matches recall by interpolating measured fixed-probe settings. Even then, \sys{} remains $1.09$--$1.22\times$ faster across all four workloads.

\subsection{Why It Is Faster}\label{section_eval_mechanism}

\begin{figure}[t]
    \centering
    \includegraphics[width=3.34in]{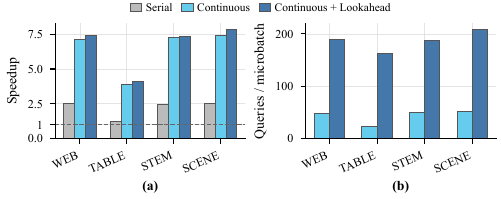}
    \figcaption{Execution-mechanism ablation at $W=8$.}{Wall time is uninstrumented and counters are diagnostic.}
    \label{fig:execution-mechanism}
    \Description{Grouped bar charts compare serial traversal, continuous batching, and continuous batching with lookahead, then compare mean queries per microbatch with and without lookahead across four workloads.}
    \vspace{-3pt}
\end{figure}

\noindent\textbf{Execution Ablation.}
Reducing partition scans is useful only if the remaining work still executes efficiently. Figure~\ref{fig:execution-mechanism}(a) separates work skipping, continuous batching, and lookahead at $W=8$. Serial traversal applies the stopping rule but fragments the workload into small partition calls, reaching only $1.25$--$2.54\times$ speedup. Continuous batching regroups ready queries by partition and raises the speedup to $3.89$--$7.40\times$. Lookahead then exposes committed ranks earlier and reaches $4.12$--$7.86\times$. It is most useful for small request batches because it parallelizes committed partition searches within each query, as Figure~\ref{fig:robustness}(a) shows. Figure~\ref{fig:execution-mechanism}(b) shows the corresponding increase in mean microbatch occupancy, from 24--51 queries without lookahead to 162--209 with it. These larger microbatches also reduce loads into Lance's partition cache by $2.7$--$3.4\times$. Because lookahead leaves the committed searches unchanged, its additional gain therefore comes entirely from fuller microbatches.

\begin{figure}[t]
    \centering
    \includegraphics[width=3.34in,trim=0 1pt 0 2pt,clip]{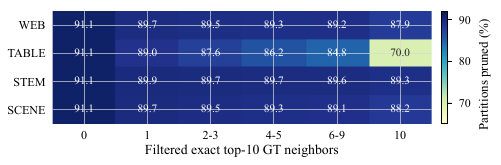}
    \figcaption{Pruned probes by threshold-neighbor count.}{}
    \label{fig:query-adaptivity}
    \Description{A heat map shows the percentage of partitions pruned for six filtered ground-truth-count groups across WEB, TABLE, STEM, and SCENE.}
\end{figure}

\noindent\textbf{Per-Query Adaptivity.}
Fixed probing visits 90 partitions per query. At $W=8$, \sys{} visits only 8.16--17.05 on average, reducing total partition visits by 81.05--90.94\%. This reduction is not limited to queries with no qualifying neighbor. Figure~\ref{fig:query-adaptivity} groups queries by their filtered exact top-10 neighbor count. Even queries with ten such neighbors avoid 70.0\% of partition visits on TABLE and 87.9--89.3\% on the other workloads. Thus, the savings depend on how early qualifying neighbors appear in the partition order, rather than merely on how many of them each query has.

\subsection{Operating Envelope}\label{section_eval_envelope}

\begin{figure}[t]
    \centering
    \includegraphics[width=3.34in]{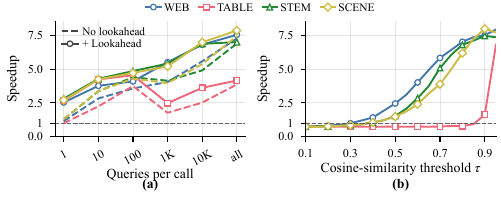}
    \figcaption{Sensitivity to batch size and threshold.}{}
    \label{fig:robustness}
    \Description{Two line charts show speedup versus queries per call and cosine-similarity threshold for four workloads. In the batch-size panel, dashed lines show continuous batching without lookahead and solid marked lines add lookahead.}
\end{figure}

\begin{figure}[t]
    \centering
    \includegraphics[width=3.34in]{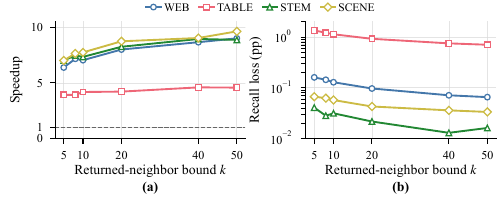}
    \figcaption{Sensitivity to the returned-neighbor bound $k$.}{}
    \label{fig:k-sensitivity}
    \Description{Two line charts show speedup and pooled filtered recall loss as the returned-neighbor bound k increases for WEB, TABLE, STEM, and SCENE.}
\end{figure}

Request batch size sets how many queries each call contains. The similarity threshold decides which candidate vectors qualify, and $k$ caps how many are returned per query. Figures~\ref{fig:robustness} and~\ref{fig:k-sensitivity} vary these parameters independently to show where \sys{} is effective.

\noindent\textbf{Request Batch Size.}
Figure~\ref{fig:robustness}(a) compares continuous batching with and without lookahead as each call grows from one query to the full 100K-query set. With one query per call, continuous batching alone reaches only $0.99$--$1.29\times$ speedup, while lookahead raises this to $2.5$--$2.8\times$ by parallelizing committed partition searches within the query. As the batch grows, continuous batching can group work from more queries, so lookahead's additional advantage narrows to $1.02$--$1.08\times$ for the full set.

\noindent\textbf{Threshold Selectivity.}
Figure~\ref{fig:robustness}(b) varies $\tau$ from 0.10 to 0.95. At loose thresholds, most partition searches return a qualifying candidate, leaving little work to prune. Scheduling overhead can then outweigh the savings, although throughput remains $0.71$--$0.99\times$ that of fixed probing in this unfavorable range. The first measured point above $1\times$ is $\tau=0.35$ for WEB, 0.90 for TABLE, and 0.40 for STEM and SCENE. In practice, semantic deduplication commonly uses high similarity thresholds~\cite{semdedup}, so \sys{} can operate in the range where it accelerates search. At $\tau=0.95$, all four workloads reach $6.8$--$7.9\times$ speedup over fixed probing.

\noindent\textbf{Returned-Neighbor Bound.}
Figure~\ref{fig:k-sensitivity} varies $k$ from 5 to 50 at each workload threshold. This bound controls how many qualifying neighbors the deduplication pipeline can retain for each query. \sys{} remains faster at every point, by $3.9$--$4.6\times$ on TABLE and $6.4$--$9.6\times$ on the other workloads, while pooled filtered recall loss stays below 1.36 percentage points. Larger $k$ increases within-partition search and merge work for both executors, but it does not remove the threshold-empty signal used for stopping.

\subsection{Impact of Approximate Assignment}\label{section_eval_assignment}

\begin{figure}[t]
    \centering
    \includegraphics[width=3.34in]{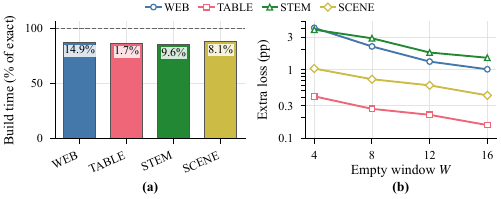}
    \figcaption{HNSW-assisted assignment at build time.}{Bar labels show nearest-centroid assignment mismatch.}
    \label{fig:assignment-precondition}
    \Description{A bar chart shows HNSW-assisted full-build time as a percentage of exact-build time, with assignment mismatch labels. A line chart shows the additional pooled filtered top-10 pruning recall loss across empty-window sizes for four workloads.}
\end{figure}

\noindent\textbf{Build-Time Assignment.}
An IVF build assigns each base vector to a centroid and stores it in the corresponding IVF partition. Exact assignment uses the nearest centroid, while Lance can approximate this step with HNSW~\cite{hnsw} to reduce centroid-distance computations. Because \sys{} follows the query-time centroid order, a vector stored under a different centroid may be visited later. We compare exact and HNSW-assisted builds with the same base vectors and centroids. Figure~\ref{fig:assignment-precondition}(a) shows that HNSW shortens full-build wall time by 12.1--14.8\%, but assigns 1.73--14.89\% of sampled vectors to a non-nearest centroid. Exact assignment shows no mismatch.

\noindent\textbf{Recall Impact.}
Figure~\ref{fig:assignment-precondition}(b) shows the consequence for early stopping. Relative to each index's own fixed-probe recall, HNSW-assisted assignment adds 0.27--2.88 percentage points of pooled recall loss at $W=8$. Increasing $W$ reduces but does not remove the gap. At $W=16$, WEB and STEM still lose an additional 1.02 and 1.50 points. We therefore disable HNSW-assisted assignment in all other experiments. Exact assignment makes each index build roughly $1.2\times$ slower, but this cost is paid once and amortized as the index is reused for daily deduplication of millions of new vectors.

\subsection{Comparison with DiskJoin}\label{section_eval_alternative}

\begin{figure}[t]
    \centering
    \includegraphics[width=3.34in]{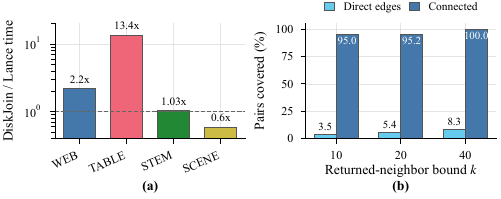}
    \figcaption{Lance and DiskJoin as different deduplication paths.}{SCENE coverage uses sampled DiskJoin pairs.}
    \label{fig:alternative-path}
    \Description{A log-scale bar chart compares DiskJoin to Lance wall time across four workloads. Grouped bars for SCENE compare direct-edge coverage with coverage after connected-components clustering at k values of 10, 20, and 40.}
\end{figure}

Our pipeline and DiskJoin~\cite{diskjoin} represent two different deduplication paths. DiskJoin targets the threshold-qualified pair relation, whereas our pipeline retains at most $k$ qualifying neighbors per query and forms duplicate groups through connected components. Figure~\ref{fig:alternative-path}(a) compares their wall time for a full self-search in which all 9.9M base vectors query the same base. For a conservative comparison, we exclude the cost of materializing DiskJoin's complete pair output. With $k=10$, Lance with \sys{} finishes $2.2\times$ and $13.4\times$ sooner on WEB and TABLE. DiskJoin is particularly slow on TABLE because the workload's embeddings are more densely clustered, causing it to evaluate substantially more candidate pairs. The two paths differ by only 3.1\% on STEM, while DiskJoin finishes $1.7\times$ sooner on SCENE. These mixed results show that neither path consistently dominates across workloads and output requirements.

Figure~\ref{fig:alternative-path}(b) asks whether returning only $k$ neighbors per query is still sufficient for deduplication. For 990 uniformly spaced SCENE samples, only 3.5--8.3\% of the pairs found by DiskJoin appear as direct Lance edges across $k=10$, 20, and 40. After connected components, however, the two records are grouped together for 95.0--100\% of these pairs. In other words, for almost every sampled pair that DiskJoin identifies as duplicate, our pipeline also groups the two records together while retaining far fewer edges. The two paths also suit different data lifecycles. DiskJoin is attractive for one-time deduplication of a largely static corpus. In our setting, millions of new vectors arrive each day, so the indexed path amortizes its build cost by reusing the same index for each arriving batch. This recurring workflow motivates our use of \sys{}.

\section{Related Work}\label{section_related_work}

\noindent\textbf{Deduplication and Threshold Joins.}
Semantic deduplication methods such as SemDeDup use pretrained embeddings to identify redundant training examples and study removal policies~\cite{semdedup}. Classical pipelines rely on keys, shingles, MinHash, or SimHash~\cite{broder1997resemblance,charikar2002similarity}. Similarity joins target threshold-qualified vector pairs. DiskJoin schedules bucket-pair work and SSD access for large self-joins~\cite{diskjoin}, while recent approximate vector-join work shares graph traversal across related queries~\cite{kim2026vectorjoin}. Unlike these pair-oriented systems, \sys{} accelerates bounded top-$k$ threshold retrieval and leaves deduplication policy downstream.

\noindent\textbf{Adaptive Vector Search and Batching.}
Vector systems increasingly incorporate predicates and bounds into search~\cite{tribase,trim}. VBASE composes vector and relational operators through relaxed monotonicity~\cite{vbase}, while Filtered-DiskANN builds label-aware graph indexes~\cite{filtereddiskann}. Existing approaches learn stopping decisions from intermediate features~\cite{li2020learned}, derive probabilistic signals from locality-sensitive hashing~\cite{jiang2025adaptive}, or use a learned recall predictor during HNSW or IVF traversal~\cite{darth}. \sys{} instead stops from observed runs of threshold-empty partitions. It needs neither training queries nor a predictor and exposes $W$ as an auditable speed/recall knob.

Database engines have long adapted execution after observing intermediate results, including Eddies and threshold algorithms~\cite{avnur2000eddies,fagin2003optimal}. \sys{} applies this idea inside an IVF physical operator, where adaptation must preserve partition locality. Faiss partition-major batching forms query--partition groups from each query's fixed $nprobe$ set before execution~\cite{faiss}. \sys{} makes group membership data-dependent. A query enters a group only when its stopping state admits the next rank, while lookahead exposes only ranks already committed by the rule. Its continuous-batching terminology follows neural model serving~\cite{orca,vllm}, but it batches queries ready for the same IVF partition rather than decoding steps.

\section{Conclusion}\label{section_conclusion}

In training data deduplication, fixed-probe IVF gives every query the same partition budget. \sys{} converts runs of threshold-empty partitions into per-query stopping signals and uses continuous batching and lookahead to preserve efficient partition-major execution. At $W=8$, \sys{} is $4.1$--$7.6\times$ faster than fixed-probe IVF on four 10M Hunyuan workloads and $6.1$--$8.4\times$ faster on two public 100M workloads under the same index and search parameters, with pooled filtered top-10 recall losses of 0.03--1.13 percentage points on Hunyuan and 1.43--2.29 percentage points on the public workloads. \sys{} accelerates IVF candidate retrieval rather than the end-to-end deduplication pipeline. More broadly, \sys{} shows how a predicate known before search can guide physical work rather than serve only as an output filter, without changing either the index format or the existing bounded top-$k$ interface.

\bibliographystyle{ACM-Reference-Format}
\bibliography{SieveIVF}


\begin{thebibliography}{26}


\ifx \showCODEN    \undefined \def \showCODEN     #1{\unskip}     \fi
\ifx \showDOI      \undefined \def \showDOI       #1{#1}\fi
\ifx \showISBNx    \undefined \def \showISBNx     #1{\unskip}     \fi
\ifx \showISBNxiii \undefined \def \showISBNxiii  #1{\unskip}     \fi
\ifx \showISSN     \undefined \def \showISSN      #1{\unskip}     \fi
\ifx \showLCCN     \undefined \def \showLCCN      #1{\unskip}     \fi
\ifx \shownote     \undefined \def \shownote      #1{#1}          \fi
\ifx \showarticletitle \undefined \def \showarticletitle #1{#1}   \fi
\ifx \showURL      \undefined \def \showURL       {\relax}        \fi
\providecommand\bibfield[2]{#2}
\providecommand\bibinfo[2]{#2}
\providecommand\natexlab[1]{#1}
\providecommand\showeprint[2][]{arXiv:#2}

\bibitem[\protect\citeauthoryear{Abbas, Tirumala, Simig, Ganguli, and
  Morcos}{Abbas et~al\mbox{.}}{2023}]%
        {semdedup}
\bibfield{author}{\bibinfo{person}{Amro Abbas}, \bibinfo{person}{Kushal
  Tirumala}, \bibinfo{person}{Dániel Simig}, \bibinfo{person}{Surya Ganguli},
  {and} \bibinfo{person}{Ari~S. Morcos}.} \bibinfo{year}{2023}\natexlab{}.
\newblock \bibinfo{title}{SemDeDup: Data-efficient learning at web-scale
  through semantic deduplication}.
\newblock
\newblock
\showeprint[arxiv]{2303.09540}~[cs.LG]
\urldef\tempurl%
\url{https://arxiv.org/abs/2303.09540}
\showURL{%
\tempurl}


\bibitem[\protect\citeauthoryear{Andr\'{e}, Kermarrec, and
  Le~Scouarnec}{Andr\'{e} et~al\mbox{.}}{2015}]%
        {quickadc}
\bibfield{author}{\bibinfo{person}{Fabien Andr\'{e}},
  \bibinfo{person}{Anne-Marie Kermarrec}, {and} \bibinfo{person}{Nicolas
  Le~Scouarnec}.} \bibinfo{year}{2015}\natexlab{}.
\newblock \showarticletitle{Cache locality is not enough: high-performance
  nearest neighbor search with product quantization fast scan}.
\newblock \bibinfo{journal}{\emph{Proc. VLDB Endow.}} \bibinfo{volume}{9},
  \bibinfo{number}{4} (\bibinfo{date}{Dec.} \bibinfo{year}{2015}),
  \bibinfo{pages}{288–299}.
\newblock
\showISSN{2150-8097}
\urldef\tempurl%
\url{https://doi.org/10.14778/2856318.2856324}
\showDOI{\tempurl}


\bibitem[\protect\citeauthoryear{Avnur and Hellerstein}{Avnur and
  Hellerstein}{2000}]%
        {avnur2000eddies}
\bibfield{author}{\bibinfo{person}{Ron Avnur} {and} \bibinfo{person}{Joseph~M.
  Hellerstein}.} \bibinfo{year}{2000}\natexlab{}.
\newblock \showarticletitle{Eddies: continuously adaptive query processing}. In
  \bibinfo{booktitle}{\emph{Proceedings of the 2000 ACM SIGMOD International
  Conference on Management of Data}} (Dallas, Texas, USA)
  \emph{(\bibinfo{series}{SIGMOD '00})}. \bibinfo{publisher}{Association for
  Computing Machinery}, \bibinfo{address}{New York, NY, USA},
  \bibinfo{pages}{261–272}.
\newblock
\showISBNx{1581132174}
\urldef\tempurl%
\url{https://doi.org/10.1145/342009.335420}
\showDOI{\tempurl}


\bibitem[\protect\citeauthoryear{Broder}{Broder}{1997}]%
        {broder1997resemblance}
\bibfield{author}{\bibinfo{person}{A.Z. Broder}.}
  \bibinfo{year}{1997}\natexlab{}.
\newblock \showarticletitle{On the resemblance and containment of documents}.
  In \bibinfo{booktitle}{\emph{Proceedings. Compression and Complexity of
  SEQUENCES 1997 (Cat. No.97TB100171)}}. \bibinfo{pages}{21--29}.
\newblock
\urldef\tempurl%
\url{https://doi.org/10.1109/SEQUEN.1997.666900}
\showDOI{\tempurl}


\bibitem[\protect\citeauthoryear{Charikar}{Charikar}{2002}]%
        {charikar2002similarity}
\bibfield{author}{\bibinfo{person}{Moses~S. Charikar}.}
  \bibinfo{year}{2002}\natexlab{}.
\newblock \showarticletitle{Similarity estimation techniques from rounding
  algorithms}. In \bibinfo{booktitle}{\emph{Proceedings of the Thiry-Fourth
  Annual ACM Symposium on Theory of Computing}} (Montreal, Quebec, Canada)
  \emph{(\bibinfo{series}{STOC '02})}. \bibinfo{publisher}{Association for
  Computing Machinery}, \bibinfo{address}{New York, NY, USA},
  \bibinfo{pages}{380–388}.
\newblock
\showISBNx{1581134959}
\urldef\tempurl%
\url{https://doi.org/10.1145/509907.509965}
\showDOI{\tempurl}


\bibitem[\protect\citeauthoryear{Chatzakis, Papakonstantinou, and
  Palpanas}{Chatzakis et~al\mbox{.}}{2025}]%
        {darth}
\bibfield{author}{\bibinfo{person}{Manos Chatzakis}, \bibinfo{person}{Yannis
  Papakonstantinou}, {and} \bibinfo{person}{Themis Palpanas}.}
  \bibinfo{year}{2025}\natexlab{}.
\newblock \showarticletitle{DARTH: Declarative Recall Through Early Termination
  for Approximate Nearest Neighbor Search}.
\newblock \bibinfo{journal}{\emph{Proc. ACM Manag. Data}} \bibinfo{volume}{3},
  \bibinfo{number}{4}, Article \bibinfo{articleno}{242} (\bibinfo{date}{Sept.}
  \bibinfo{year}{2025}), \bibinfo{numpages}{26}~pages.
\newblock
\urldef\tempurl%
\url{https://doi.org/10.1145/3749160}
\showDOI{\tempurl}


\bibitem[\protect\citeauthoryear{Chen, Zhao, Wang, Li, Liu, Li, Yang, and
  Wang}{Chen et~al\mbox{.}}{2021}]%
        {spann}
\bibfield{author}{\bibinfo{person}{Qi Chen}, \bibinfo{person}{Bing Zhao},
  \bibinfo{person}{Haidong Wang}, \bibinfo{person}{Mingqin Li},
  \bibinfo{person}{Chuanjie Liu}, \bibinfo{person}{Zengzhong Li},
  \bibinfo{person}{Mao Yang}, {and} \bibinfo{person}{Jingdong Wang}.}
  \bibinfo{year}{2021}\natexlab{}.
\newblock \showarticletitle{SPANN: highly-efficient billion-scale approximate
  nearest neighbor search}. In \bibinfo{booktitle}{\emph{Proceedings of the
  35th International Conference on Neural Information Processing Systems}}
  \emph{(\bibinfo{series}{NIPS '21})}. \bibinfo{publisher}{Curran Associates
  Inc.}, \bibinfo{address}{Red Hook, NY, USA}, Article
  \bibinfo{articleno}{398}, \bibinfo{numpages}{14}~pages.
\newblock
\showISBNx{9781713845393}


\bibitem[\protect\citeauthoryear{Chen, Yan, Meliou, and Lo}{Chen
  et~al\mbox{.}}{2025}]%
        {diskjoin}
\bibfield{author}{\bibinfo{person}{Yanqi Chen}, \bibinfo{person}{Xiao Yan},
  \bibinfo{person}{Alexandra Meliou}, {and} \bibinfo{person}{Eric Lo}.}
  \bibinfo{year}{2025}\natexlab{}.
\newblock \showarticletitle{DiskJoin: Large-scale Vector Similarity Join with
  SSD}.
\newblock \bibinfo{journal}{\emph{Proc. ACM Manag. Data}} \bibinfo{volume}{3},
  \bibinfo{number}{6}, Article \bibinfo{articleno}{315} (\bibinfo{date}{Dec.}
  \bibinfo{year}{2025}), \bibinfo{numpages}{27}~pages.
\newblock
\urldef\tempurl%
\url{https://doi.org/10.1145/3769780}
\showDOI{\tempurl}


\bibitem[\protect\citeauthoryear{Fagin, Lotem, and Naor}{Fagin
  et~al\mbox{.}}{2001}]%
        {fagin2003optimal}
\bibfield{author}{\bibinfo{person}{Ronald Fagin}, \bibinfo{person}{Amnon
  Lotem}, {and} \bibinfo{person}{Moni Naor}.} \bibinfo{year}{2001}\natexlab{}.
\newblock \showarticletitle{Optimal aggregation algorithms for middleware}. In
  \bibinfo{booktitle}{\emph{Proceedings of the Twentieth ACM
  SIGMOD-SIGACT-SIGART Symposium on Principles of Database Systems}} (Santa
  Barbara, California, USA) \emph{(\bibinfo{series}{PODS '01})}.
  \bibinfo{publisher}{Association for Computing Machinery},
  \bibinfo{address}{New York, NY, USA}, \bibinfo{pages}{102–113}.
\newblock
\showISBNx{1581133618}
\urldef\tempurl%
\url{https://doi.org/10.1145/375551.375567}
\showDOI{\tempurl}


\bibitem[\protect\citeauthoryear{Gao and Long}{Gao and Long}{2024}]%
        {rabitq}
\bibfield{author}{\bibinfo{person}{Jianyang Gao} {and} \bibinfo{person}{Cheng
  Long}.} \bibinfo{year}{2024}\natexlab{}.
\newblock \showarticletitle{RaBitQ: Quantizing High-Dimensional Vectors with a
  Theoretical Error Bound for Approximate Nearest Neighbor Search}.
\newblock \bibinfo{journal}{\emph{Proc. ACM Manag. Data}} \bibinfo{volume}{2},
  \bibinfo{number}{3}, Article \bibinfo{articleno}{167} (\bibinfo{date}{May}
  \bibinfo{year}{2024}), \bibinfo{numpages}{27}~pages.
\newblock
\urldef\tempurl%
\url{https://doi.org/10.1145/3654970}
\showDOI{\tempurl}


\bibitem[\protect\citeauthoryear{Gollapudi, Karia, Sivashankar, Krishnaswamy,
  Begwani, Raz, Lin, Zhang, Mahapatro, Srinivasan, Singh, and
  Simhadri}{Gollapudi et~al\mbox{.}}{2023}]%
        {filtereddiskann}
\bibfield{author}{\bibinfo{person}{Siddharth Gollapudi}, \bibinfo{person}{Neel
  Karia}, \bibinfo{person}{Varun Sivashankar}, \bibinfo{person}{Ravishankar
  Krishnaswamy}, \bibinfo{person}{Nikit Begwani}, \bibinfo{person}{Swapnil
  Raz}, \bibinfo{person}{Yiyong Lin}, \bibinfo{person}{Yin Zhang},
  \bibinfo{person}{Neelam Mahapatro}, \bibinfo{person}{Premkumar Srinivasan},
  \bibinfo{person}{Amit Singh}, {and} \bibinfo{person}{Harsha~Vardhan
  Simhadri}.} \bibinfo{year}{2023}\natexlab{}.
\newblock \showarticletitle{Filtered-DiskANN: Graph Algorithms for Approximate
  Nearest Neighbor Search with Filters}. In
  \bibinfo{booktitle}{\emph{Proceedings of the ACM Web Conference 2023}}
  (Austin, TX, USA) \emph{(\bibinfo{series}{WWW '23})}.
  \bibinfo{publisher}{Association for Computing Machinery},
  \bibinfo{address}{New York, NY, USA}, \bibinfo{pages}{3406–3416}.
\newblock
\showISBNx{9781450394161}
\urldef\tempurl%
\url{https://doi.org/10.1145/3543507.3583552}
\showDOI{\tempurl}


\bibitem[\protect\citeauthoryear{Jegou, Douze, and Schmid}{Jegou
  et~al\mbox{.}}{2011}]%
        {jegou2011product}
\bibfield{author}{\bibinfo{person}{Herve Jegou}, \bibinfo{person}{Matthijs
  Douze}, {and} \bibinfo{person}{Cordelia Schmid}.}
  \bibinfo{year}{2011}\natexlab{}.
\newblock \showarticletitle{Product Quantization for Nearest Neighbor Search}.
\newblock \bibinfo{journal}{\emph{IEEE Trans. Pattern Anal. Mach. Intell.}}
  \bibinfo{volume}{33}, \bibinfo{number}{1} (\bibinfo{date}{Jan.}
  \bibinfo{year}{2011}), \bibinfo{pages}{117–128}.
\newblock
\showISSN{0162-8828}
\urldef\tempurl%
\url{https://doi.org/10.1109/TPAMI.2010.57}
\showDOI{\tempurl}


\bibitem[\protect\citeauthoryear{Jiang, Xu, and Gao}{Jiang
  et~al\mbox{.}}{2025}]%
        {jiang2025adaptive}
\bibfield{author}{\bibinfo{person}{Jianfeng Jiang}, \bibinfo{person}{Shen Xu},
  {and} \bibinfo{person}{Ying Gao}.} \bibinfo{year}{2025}\natexlab{}.
\newblock \showarticletitle{A uniformed adaptive early termination model
  through probabilistic feature to speed up quantization-based search}.
\newblock \bibinfo{journal}{\emph{International Journal of Computers and
  Applications}} \bibinfo{volume}{47}, \bibinfo{number}{1}
  (\bibinfo{year}{2025}), \bibinfo{pages}{17--28}.
\newblock
\urldef\tempurl%
\url{https://doi.org/10.1080/1206212X.2024.2431229}
\showDOI{\tempurl}


\bibitem[\protect\citeauthoryear{Johnson, Douze, and Jégou}{Johnson
  et~al\mbox{.}}{2021}]%
        {faiss}
\bibfield{author}{\bibinfo{person}{Jeff Johnson}, \bibinfo{person}{Matthijs
  Douze}, {and} \bibinfo{person}{Hervé Jégou}.}
  \bibinfo{year}{2021}\natexlab{}.
\newblock \showarticletitle{Billion-Scale Similarity Search with GPUs}.
\newblock \bibinfo{journal}{\emph{IEEE Transactions on Big Data}}
  \bibinfo{volume}{7}, \bibinfo{number}{3} (\bibinfo{year}{2021}),
  \bibinfo{pages}{535--547}.
\newblock


\bibitem[\protect\citeauthoryear{Kim, Roth, Liang, and Ailamaki}{Kim
  et~al\mbox{.}}{2026}]%
        {kim2026vectorjoin}
\bibfield{author}{\bibinfo{person}{Kyoungmin Kim}, \bibinfo{person}{Lennart
  Roth}, \bibinfo{person}{Liang Liang}, {and} \bibinfo{person}{Anastasia
  Ailamaki}.} \bibinfo{year}{2026}\natexlab{}.
\newblock \bibinfo{title}{Fast Approximate Vector Joins via Offline-Online
  Co-Design}.
\newblock
\newblock
\showeprint[arxiv]{2603.16360}~[cs.DB]
\urldef\tempurl%
\url{https://arxiv.org/abs/2603.16360}
\showURL{%
\tempurl}


\bibitem[\protect\citeauthoryear{Kwon, Li, Zhuang, Sheng, Zheng, Yu, Gonzalez,
  Zhang, and Stoica}{Kwon et~al\mbox{.}}{2023}]%
        {vllm}
\bibfield{author}{\bibinfo{person}{Woosuk Kwon}, \bibinfo{person}{Zhuohan Li},
  \bibinfo{person}{Siyuan Zhuang}, \bibinfo{person}{Ying Sheng},
  \bibinfo{person}{Lianmin Zheng}, \bibinfo{person}{Cody~Hao Yu},
  \bibinfo{person}{Joseph Gonzalez}, \bibinfo{person}{Hao Zhang}, {and}
  \bibinfo{person}{Ion Stoica}.} \bibinfo{year}{2023}\natexlab{}.
\newblock \showarticletitle{Efficient Memory Management for Large Language
  Model Serving with PagedAttention}. In \bibinfo{booktitle}{\emph{Proceedings
  of the 29th Symposium on Operating Systems Principles}} (Koblenz, Germany)
  \emph{(\bibinfo{series}{SOSP '23})}. \bibinfo{publisher}{Association for
  Computing Machinery}, \bibinfo{address}{New York, NY, USA},
  \bibinfo{pages}{611–626}.
\newblock
\showISBNx{9798400702297}
\urldef\tempurl%
\url{https://doi.org/10.1145/3600006.3613165}
\showDOI{\tempurl}


\bibitem[\protect\citeauthoryear{{Lance Format}}{{Lance Format}}{2026}]%
        {lance}
\bibfield{author}{\bibinfo{person}{{Lance Format}}.}
  \bibinfo{year}{2026}\natexlab{}.
\newblock \bibinfo{title}{Lance: The Open Lakehouse Format for Multimodal
  {AI}}.
\newblock \bibinfo{howpublished}{\url{https://github.com/lance-format/lance}}.
\newblock


\bibitem[\protect\citeauthoryear{Li, Zhang, Andersen, and He}{Li
  et~al\mbox{.}}{2020}]%
        {li2020learned}
\bibfield{author}{\bibinfo{person}{Conglong Li}, \bibinfo{person}{Minjia
  Zhang}, \bibinfo{person}{David~G. Andersen}, {and} \bibinfo{person}{Yuxiong
  He}.} \bibinfo{year}{2020}\natexlab{}.
\newblock \showarticletitle{Improving Approximate Nearest Neighbor Search
  through Learned Adaptive Early Termination}. In
  \bibinfo{booktitle}{\emph{Proceedings of the 2020 ACM SIGMOD International
  Conference on Management of Data}} (Portland, OR, USA)
  \emph{(\bibinfo{series}{SIGMOD '20})}. \bibinfo{publisher}{Association for
  Computing Machinery}, \bibinfo{address}{New York, NY, USA},
  \bibinfo{pages}{2539–2554}.
\newblock
\showISBNx{9781450367356}
\urldef\tempurl%
\url{https://doi.org/10.1145/3318464.3380600}
\showDOI{\tempurl}


\bibitem[\protect\citeauthoryear{Malkov and Yashunin}{Malkov and
  Yashunin}{2020}]%
        {hnsw}
\bibfield{author}{\bibinfo{person}{Yu~A. Malkov} {and} \bibinfo{person}{D.~A.
  Yashunin}.} \bibinfo{year}{2020}\natexlab{}.
\newblock \showarticletitle{Efficient and Robust Approximate Nearest Neighbor
  Search Using Hierarchical Navigable Small World Graphs}.
\newblock \bibinfo{journal}{\emph{IEEE Transactions on Pattern Analysis and
  Machine Intelligence}} \bibinfo{volume}{42}, \bibinfo{number}{4}
  (\bibinfo{year}{2020}), \bibinfo{pages}{824--836}.
\newblock
\urldef\tempurl%
\url{https://doi.org/10.1109/TPAMI.2018.2889473}
\showDOI{\tempurl}


\bibitem[\protect\citeauthoryear{Schuhmann, Beaumont, Vencu, Gordon, Wightman,
  Cherti, Coombes, Katta, Mullis, Wortsman, Schramowski, Kundurthy, Crowson,
  Schmidt, Kaczmarczyk, and Jitsev}{Schuhmann et~al\mbox{.}}{2022}]%
        {laion5b}
\bibfield{author}{\bibinfo{person}{Christoph Schuhmann},
  \bibinfo{person}{Romain Beaumont}, \bibinfo{person}{Richard Vencu},
  \bibinfo{person}{Cade Gordon}, \bibinfo{person}{Ross Wightman},
  \bibinfo{person}{Mehdi Cherti}, \bibinfo{person}{Theo Coombes},
  \bibinfo{person}{Aarush Katta}, \bibinfo{person}{Clayton Mullis},
  \bibinfo{person}{Mitchell Wortsman}, \bibinfo{person}{Patrick Schramowski},
  \bibinfo{person}{Srivatsa Kundurthy}, \bibinfo{person}{Katherine Crowson},
  \bibinfo{person}{Ludwig Schmidt}, \bibinfo{person}{Robert Kaczmarczyk}, {and}
  \bibinfo{person}{Jenia Jitsev}.} \bibinfo{year}{2022}\natexlab{}.
\newblock \showarticletitle{LAION-5B: an open large-scale dataset for training
  next generation image-text models}. In \bibinfo{booktitle}{\emph{Proceedings
  of the 36th International Conference on Neural Information Processing
  Systems}} (New Orleans, LA, USA) \emph{(\bibinfo{series}{NIPS '22})}.
  \bibinfo{publisher}{Curran Associates Inc.}, \bibinfo{address}{Red Hook, NY,
  USA}, Article \bibinfo{articleno}{1833}, \bibinfo{numpages}{17}~pages.
\newblock


\bibitem[\protect\citeauthoryear{Song, Zhang, Gao, Yao, Wang, Wu, and Qu}{Song
  et~al\mbox{.}}{2025}]%
        {trim}
\bibfield{author}{\bibinfo{person}{Yitong Song}, \bibinfo{person}{Pengcheng
  Zhang}, \bibinfo{person}{Chao Gao}, \bibinfo{person}{Bin Yao},
  \bibinfo{person}{Kai Wang}, \bibinfo{person}{Zongyuan Wu}, {and}
  \bibinfo{person}{Lin Qu}.} \bibinfo{year}{2025}\natexlab{}.
\newblock \showarticletitle{TRIM: Accelerating High-Dimensional Vector
  Similarity Search with Enhanced Triangle-Inequality-Based Pruning}.
\newblock \bibinfo{journal}{\emph{Proc. ACM Manag. Data}} \bibinfo{volume}{3},
  \bibinfo{number}{6}, Article \bibinfo{articleno}{373} (\bibinfo{date}{Dec.}
  \bibinfo{year}{2025}), \bibinfo{numpages}{26}~pages.
\newblock
\urldef\tempurl%
\url{https://doi.org/10.1145/3769838}
\showDOI{\tempurl}


\bibitem[\protect\citeauthoryear{Wang, Yi, Guo, Jin, Xu, Li, Wang, Guo, Li, Xu,
  Yu, Yuan, Zou, Long, Cai, Li, Zhang, Mo, Gu, Jiang, Wei, and Xie}{Wang
  et~al\mbox{.}}{2021}]%
        {milvus}
\bibfield{author}{\bibinfo{person}{Jianguo Wang}, \bibinfo{person}{Xiaomeng
  Yi}, \bibinfo{person}{Rentong Guo}, \bibinfo{person}{Hai Jin},
  \bibinfo{person}{Peng Xu}, \bibinfo{person}{Shengjun Li},
  \bibinfo{person}{Xiangyu Wang}, \bibinfo{person}{Xiangzhou Guo},
  \bibinfo{person}{Chengming Li}, \bibinfo{person}{Xiaohai Xu},
  \bibinfo{person}{Kun Yu}, \bibinfo{person}{Yuxing Yuan},
  \bibinfo{person}{Yinghao Zou}, \bibinfo{person}{Jiquan Long},
  \bibinfo{person}{Yudong Cai}, \bibinfo{person}{Zhenxiang Li},
  \bibinfo{person}{Zhifeng Zhang}, \bibinfo{person}{Yihua Mo},
  \bibinfo{person}{Jun Gu}, \bibinfo{person}{Ruiyi Jiang}, \bibinfo{person}{Yi
  Wei}, {and} \bibinfo{person}{Charles Xie}.} \bibinfo{year}{2021}\natexlab{}.
\newblock \showarticletitle{Milvus: A Purpose-Built Vector Data Management
  System}. In \bibinfo{booktitle}{\emph{Proceedings of the 2021 International
  Conference on Management of Data}} (Virtual Event, China)
  \emph{(\bibinfo{series}{SIGMOD '21})}. \bibinfo{publisher}{Association for
  Computing Machinery}, \bibinfo{address}{New York, NY, USA},
  \bibinfo{pages}{2614–2627}.
\newblock
\showISBNx{9781450383431}
\urldef\tempurl%
\url{https://doi.org/10.1145/3448016.3457550}
\showDOI{\tempurl}


\bibitem[\protect\citeauthoryear{Xu, Yang, Zhang, Pan, Chen, Shen, Zhou, and
  Du}{Xu et~al\mbox{.}}{2025}]%
        {tribase}
\bibfield{author}{\bibinfo{person}{Qian Xu}, \bibinfo{person}{Juan Yang},
  \bibinfo{person}{Feng Zhang}, \bibinfo{person}{Junda Pan},
  \bibinfo{person}{Kang Chen}, \bibinfo{person}{Youren Shen},
  \bibinfo{person}{Amelie~Chi Zhou}, {and} \bibinfo{person}{Xiaoyong Du}.}
  \bibinfo{year}{2025}\natexlab{}.
\newblock \showarticletitle{Tribase: A Vector Data Query Engine for Reliable
  and Lossless Pruning Compression using Triangle Inequalities}.
\newblock \bibinfo{journal}{\emph{Proc. ACM Manag. Data}} \bibinfo{volume}{3},
  \bibinfo{number}{1}, Article \bibinfo{articleno}{82} (\bibinfo{date}{Feb.}
  \bibinfo{year}{2025}), \bibinfo{numpages}{28}~pages.
\newblock
\urldef\tempurl%
\url{https://doi.org/10.1145/3709743}
\showDOI{\tempurl}


\bibitem[\protect\citeauthoryear{Yandex and Lempitsky}{Yandex and
  Lempitsky}{2016}]%
        {deep1b}
\bibfield{author}{\bibinfo{person}{Artem~Babenko Yandex} {and}
  \bibinfo{person}{Victor Lempitsky}.} \bibinfo{year}{2016}\natexlab{}.
\newblock \showarticletitle{Efficient Indexing of Billion-Scale Datasets of
  Deep Descriptors}. In \bibinfo{booktitle}{\emph{2016 IEEE Conference on
  Computer Vision and Pattern Recognition (CVPR)}}.
  \bibinfo{pages}{2055--2063}.
\newblock
\urldef\tempurl%
\url{https://doi.org/10.1109/CVPR.2016.226}
\showDOI{\tempurl}


\bibitem[\protect\citeauthoryear{Yu, Jeong, Kim, Kim, and Chun}{Yu
  et~al\mbox{.}}{2022}]%
        {orca}
\bibfield{author}{\bibinfo{person}{Gyeong-In Yu}, \bibinfo{person}{Joo~Seong
  Jeong}, \bibinfo{person}{Geon-Woo Kim}, \bibinfo{person}{Soojeong Kim}, {and}
  \bibinfo{person}{Byung-Gon Chun}.} \bibinfo{year}{2022}\natexlab{}.
\newblock \showarticletitle{Orca: A Distributed Serving System for
  {Transformer-Based} Generative Models}. In \bibinfo{booktitle}{\emph{16th
  USENIX Symposium on Operating Systems Design and Implementation (OSDI 22)}}.
  \bibinfo{publisher}{USENIX Association}, \bibinfo{address}{Carlsbad, CA},
  \bibinfo{pages}{521--538}.
\newblock
\showISBNx{978-1-939133-28-1}


\bibitem[\protect\citeauthoryear{Zhang, Xu, Chen, Sui, Xie, Cai, Chen, He,
  Yang, Yang, Yang, and Zhou}{Zhang et~al\mbox{.}}{2023}]%
        {vbase}
\bibfield{author}{\bibinfo{person}{Qianxi Zhang}, \bibinfo{person}{Shuotao Xu},
  \bibinfo{person}{Qi Chen}, \bibinfo{person}{Guoxin Sui},
  \bibinfo{person}{Jiadong Xie}, \bibinfo{person}{Zhizhen Cai},
  \bibinfo{person}{Yaoqi Chen}, \bibinfo{person}{Yinxuan He},
  \bibinfo{person}{Yuqing Yang}, \bibinfo{person}{Fan Yang},
  \bibinfo{person}{Mao Yang}, {and} \bibinfo{person}{Lidong Zhou}.}
  \bibinfo{year}{2023}\natexlab{}.
\newblock \showarticletitle{{VBASE}: Unifying Online Vector Similarity Search
  and Relational Queries via Relaxed Monotonicity}. In
  \bibinfo{booktitle}{\emph{17th USENIX Symposium on Operating Systems Design
  and Implementation (OSDI 23)}}. \bibinfo{publisher}{USENIX Association},
  \bibinfo{address}{Boston, MA}, \bibinfo{pages}{377--395}.
\newblock
\showISBNx{978-1-939133-34-2}


\end{thebibliography}

\appendix
\section{Automatic Window Calibration}\label{section_appendix_calibration}

The consecutive-empty window $W$ gives \sys{} a single, transparent speed/recall knob, but the best setting varies across workloads and even across independently indexed shards. Selecting it offline would require profiling every new workload, potentially against exact ground truth. We therefore add a calibrated execution mode that selects $W$ automatically inside each sufficiently large search call. It uses fixed-probe output from a small sample of the same query batch and requires neither exact ground truth nor an offline workload profile. Sampling works because a call fixes the query distribution, threshold, and index, so a uniform sample observes the same mix of stopping behavior as the full batch.

\noindent\textbf{In-Call Calibration.}
Given a query batch $Q$, the executor uniformly samples $Q_s\subset Q$ without replacement and fully probes its $nprobe$ partitions. It retains the resulting partition-level outputs and merges them into the fixed-probe result $A_{\mathrm{fixed}}(q)$ for every sampled query. The executor then replays the stopping rule over these recorded outcomes for each candidate $W$. Let $A_W(q)$ be the output that window $W$ would retain. We measure its loss relative to the sampled fixed-probe output as
\[
\operatorname{loss}(W)=1-
\frac{\sum_{q\in Q_s}|A_W(q)\cap A_{\mathrm{fixed}}(q)|}
     {\sum_{q\in Q_s}|A_{\mathrm{fixed}}(q)|}.
\]
The executor selects the smallest $W<nprobe$ satisfying \(\operatorname{loss}(W)<\epsilon\), where \(\epsilon\) is the user-specified loss target. If no pruned window satisfies the target, it uses full probing. The sampled queries keep their already computed fixed-probe results, while the remaining queries run with lookahead under the selected $W$.

\begin{algorithm}[H]
\caption{Automatic consecutive-empty window calibration}
\label{alg:calibrated-window}
\begin{SievePseudocode}
\begin{algorithmic}[1]
\State \AlgNote{$Q$: query batch; $P$: per-query partition orders}
\State \AlgNote{$r$: sample rate; $\epsilon$: baseline-relative loss target}
\Procedure{CalibratedSearch}{$Q,P,\tau,k,nprobe,r,\epsilon$}
    \State $Q_s \gets$ \Call{UniformSample}{$Q,\lceil r|Q|\rceil$}
    \State $L_s \gets$ \Call{FullProbeByPartition}{$Q_s,P,\tau,k,nprobe$}
    \State $A_{\mathrm{fixed}} \gets$ \Call{MergeTopK}{$L_s,k$}
    \State $H \gets$ \Call{RequiredWindowHistogram}{$L_s,A_{\mathrm{fixed}}$}
    \State $Z \gets \sum_{q\in Q_s}|A_{\mathrm{fixed}}(q)|$
    \State $W^* \gets nprobe$; $retained \gets 0$
    \For{$W \gets 1$ to $nprobe-1$}
        \State $retained \gets retained + H[W]$
        \State $loss \gets 1-retained/Z$
        \If{$loss < \epsilon$}
            \State $W^* \gets W$; \AlgKeyword{break}
        \EndIf
    \EndFor
    \State $A_r \gets$ \Call{LookaheadSearch}{$Q\setminus Q_s,P,\tau,W^*,k$}
    \State \Return \Call{RestoreOrder}{$A_{\mathrm{fixed}},A_r$}
\EndProcedure
\end{algorithmic}
\end{SievePseudocode}
\end{algorithm}

\noindent\textbf{Efficient Replay.}
Calibration does not issue another search for every candidate window $W$. A lookahead execution always visits a prefix of the partition order. Cutting the probe order short only removes candidates, and a neighbor drops out of the top-$k$ only when $k$ better candidates displace it. The prefix cannot supply a better candidate that was not already there, so every neighbor of $A_{\mathrm{fixed}}(q)$ found within the prefix also appears in $A_W(q)$. Each neighbor therefore has a \emph{required window}: the smallest $W$ whose prefix reaches the partition where that neighbor was found. One pass over a sampled query's outputs, in rank order, computes it for all of them. Counting neighbors by required window gives the histogram $H$ of Algorithm~\ref{alg:calibrated-window}: pooled over all sampled queries, $H[W]$ is the number of neighbors whose required window is exactly $W$. A larger $W$ only extends the prefix, so window $W$ retains every neighbor whose required window does not exceed $W$. The prefix sums of $H$ therefore give $\operatorname{loss}(W)$ for every window at once, at $O(nprobe\cdot k)$ cost per sampled query. In our evaluated configuration, calls below 10K queries or samples that return no neighbors use fallback $W=8$. Calls without a similarity threshold use full probing because they provide no threshold-empty stopping signal.

\begin{figure}[H]
    \centering
    \includegraphics[width=3.34in]{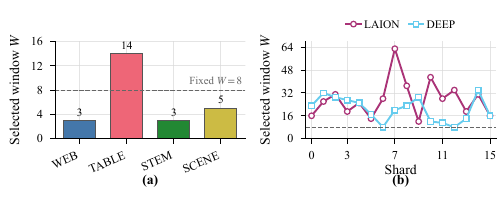}
    \figcaption{Online window selection across workloads and shards.}{Panel (a) shows one 10M index per workload. Panel (b) shows independent choices for the 16 shards of each public 100M workload.}
    \label{fig:calibrated-window}
    \Description{Panel (a) is a bar chart showing selected windows 3, 14, 3, and 5 for WEB, TABLE, STEM, and SCENE. Panel (b) plots independently selected windows across the 16 LAION and DEEP shards, ranging from 12 to 63 for LAION and from 8 to 34 for DEEP.}
\end{figure}

\noindent\textbf{Results.}
We evaluate the mechanism with a 2\% sample, a strict 0.5\% sample-relative loss target, and the same 100K-query calls described in Section~\ref{section_eval_setup}. Figure~\ref{fig:calibrated-window}(a) shows that calibration selects $W=3$, 14, 3, and 5 for WEB, TABLE, STEM, and SCENE. For the scale-out workloads, each independently indexed shard calibrates against its own local output while using the same sampled query indices. The selected windows span 12--63 across LAION's 16 shards and 8--34 across DEEP's (Figure~\ref{fig:calibrated-window}(b)). This variation confirms that one fixed window need not fit even every shard of the same workload, consistent with the view that the selected window is determined jointly by the query batch and the index it searches.

\begin{figure}[H]
    \centering
    \includegraphics[width=3.34in]{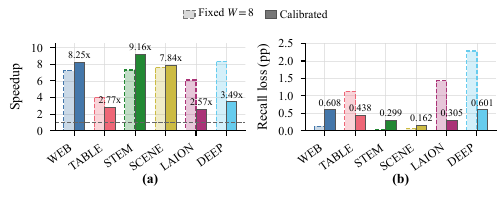}
    \figcaption{Performance and recall under fixed and calibrated windows.}{Light dashed bars use fixed $W=8$, and solid bars use automatic calibration. Panels report speedup over fixed probing and recall loss against pooled filtered exact top-10 ground truth.}
    \label{fig:calibrated-outcomes}
    \Description{Grouped bar charts compare shared W equal to 8 with automatic calibration on WEB, TABLE, STEM, SCENE, LAION-100M, and DEEP-100M. Panel (a) shows speedup over fixed probing. Panel (b) shows exact-ground-truth pooled filtered top-10 recall loss.}
\end{figure}

Figure~\ref{fig:calibrated-outcomes} compares calibration with the fixed $W=8$ setting. Including the full-probe sample cost, calibrated execution is $2.77$--$9.16\times$ faster on the 10M workloads and $2.57$--$3.49\times$ faster on the public 100M workloads than fixed probing under the same setup. It chooses more conservative windows for TABLE and both public workloads, reducing the largest exact-GT recall loss from 2.29 percentage points under $W=8$ to 0.608. Although the sample-relative calibration loss and exact-GT recall loss are different metrics, all six calibrated losses stay around or below 0.5 percentage points (range: $0.162$--$0.608$). The goal is automatic $W$ selection under an explicit loss target, not uniform speed gains over a manually chosen $W$.

\noindent\textbf{Scope.}
The 0.5\% target estimates loss against fixed-probe output on the sampled queries. It is neither a confidence bound nor a direct guarantee on full-call exact-GT recall. Sample queries receive full probing, whose cost is included in the reported wall time.

\end{document}